# Contributions of GRBs and Cen A-like Radio Galaxies to the Cosmic Gamma-ray Background

K. Watanabe[1] & D.H.Hartmann[2]

1,EIT/LHEA,NASA/GSFC, Code 660, Greenbelt, MD 20771, USA
2,Clemson University, Clemson, SC 29634-0978, USA

**Abstract.** The contribution to the cosmic diffuse gamma-ray background (CGB) from Gamma Ray Bursts (GRBs) is studied in the 40 keV - 2 MeV regime. We use High Energy Resolution (HER) data from the Burst And Transient Source Experiment (BATSE) aboard the Compton Gamma-Ray Observatory (CGRO) to generate a GRB template spectrum. Although the GRB contribution to the CGB is generally small, in comparison to the dominant flux from Type Ia supernovae, the integrated GRB flux is in fact comparable to that from SNIa in the narrow 10-40 keV range. GRBs contribute to the CGB at the same level as Type II supernovae do. Although BATSE data are not available below ~40 keV, extrapolation of the template spectrum suggests that bursts can fill a significant part of the existing gap between Seyfert galaxies (dominating the CGB below ~ 100 keV) and SNIa (dominating at ~1 MeV). We estimate contributions from Cen A-like (FR I) radio galaxies in this energy regime, where INTEGRAL data is expected to provide major advances.

## INTRODUCTION

We have studied the CGB in the MeV region both observationally and theoretically (see e.g., Watanabe et al. 1999a,b). Recent observations with COMPTEL and SMM show no evidence for a "MeV bump" (Weidenspointner 1999, Watanabe et al. 1999b) which was indicated by previous experiments. Suggested sources for the CGB include unresolved Seyfert galaxies (e.g., Zdziarski 1996) in the lower energy band (up to 300, keV) and Blazars (e.g., Sreekumar et al. 1998) in the high energy band (> 100 MeV). Although cosmological SNIa contribute a significant flux to the CGB around 1 MeV (The et al. 1993, Watanabe et al. 1998 & 1999a, Ruiz-Lapuente et al. 2001), there remains a gap between Seyfert and SNIa contributions where a continuous power-law like CGB spectrum has been observed. There is no generally accepted explanation for the origin of the CGB flux in this gap. Here we consider the contribution from GRBs and Cen A- like (FR I) radio galaxies, and show that neither of these sources can adequately fill the gap.

# DATA ANALYSIS

## Gamma-ray Bursts (GRBs)

We use the BATSE GRB trigger data from the 4B catalog, which is available online from http://cossc.gsfc.nasa.gov/cossc/batse/4Bcatalog/. For 1234 bursts in the catalog duration (T90) information is available. Using the FTOOLS CGRO sub-package and XSPEC, GRB energy spectra with background subtraction are constructed. High Energy Resolution (HER) data obtained from the detector which received the brightest gamma-ray flux among BATSE's eight detectors were selected for spectral analysis. In some cases HER data are not available or the burst is too faint to allow high quality spectral analysis. In the end, 781 spectra out of 1234 triggered events were generated. We sum all spectra (multiplied by T90) to obtain an average fluence template in units of *photons/cm$^2$/keV*. In order to obtain the GRB contribution to the CGB, the template was divided by 4π and multiplied by an average all sky GRB rate of one per day.

## Cen A-like Radio Galaxies

Centaurus A (Cen A) is a radio-loud Seyfert galaxy, belonging to a class intermediate between radio-quiet (normal) Seyferts and Blazars, also classified as FR I radio galaxy. Cen A is the brightest radio galaxy (with a Seyfert 2 nucleus) detected with OSSE, the Oriented Scintillation Spectrometer Experiment on CGRO. The spectrum of Cen A appears to extend well into the GeV regime (Kinzer et al. 1995).

We use the BATSE occultation data for Cen A available online from http://cossc.gsfc.nasa.gov/cossc/batse/hilev/CEN_A/cen_a.html/. An average Cen A spectrum was obtained from 385 individual spectra in the energy range of 20 keV < E < 1 MeV by using the FTOOLS CGRO sub-package and XSPEC. We use the average spectrum as template for all the Cen A-like (FR I) radio galaxies. We fit the spectrum with a power law with index of 1.7, and extrapolate to 10 MeV. We adopt a present-day FRI galaxy density of $n_o = 2.0 \times 10^{-6}$ galaxies/Mpc$^3$ (e.g., Canosa, et al. 1999, Colina, et al., 1995, & Colla, et al., 1975). The contribution of the Cen A-like (FR I) radio galaxies to the CGB in units of (photons cm$^2$ s$^{-1}$ keV$^{-1}$ sr$^{-1}$) is given by

$$F_e = L_H D^2 \int dz\, n(z) \dot{N}\left(E_\gamma \times (1+z)\right) E(z)$$

where n(z) accounts for density evolution: n(z)=n$_0$(1+z)$^m$ with m = 3, 2, 1, 0, -1, -2, -3, $L_H$=c/$H_0$ is the Hubble length, D is the distance to Cen A (3.4 Mpc), the dotted N is the observed Cen A spectrum (photons cm$^2$ s$^{-1}$ keV$^{-1}$ ), and where E(z) represents the evolution of the Hubble constant (see eq. [13.3] in Peebles 1993). The integration over red shift was carried out to z = 3. FIGURE 1 shows the resulting CGB contributions.

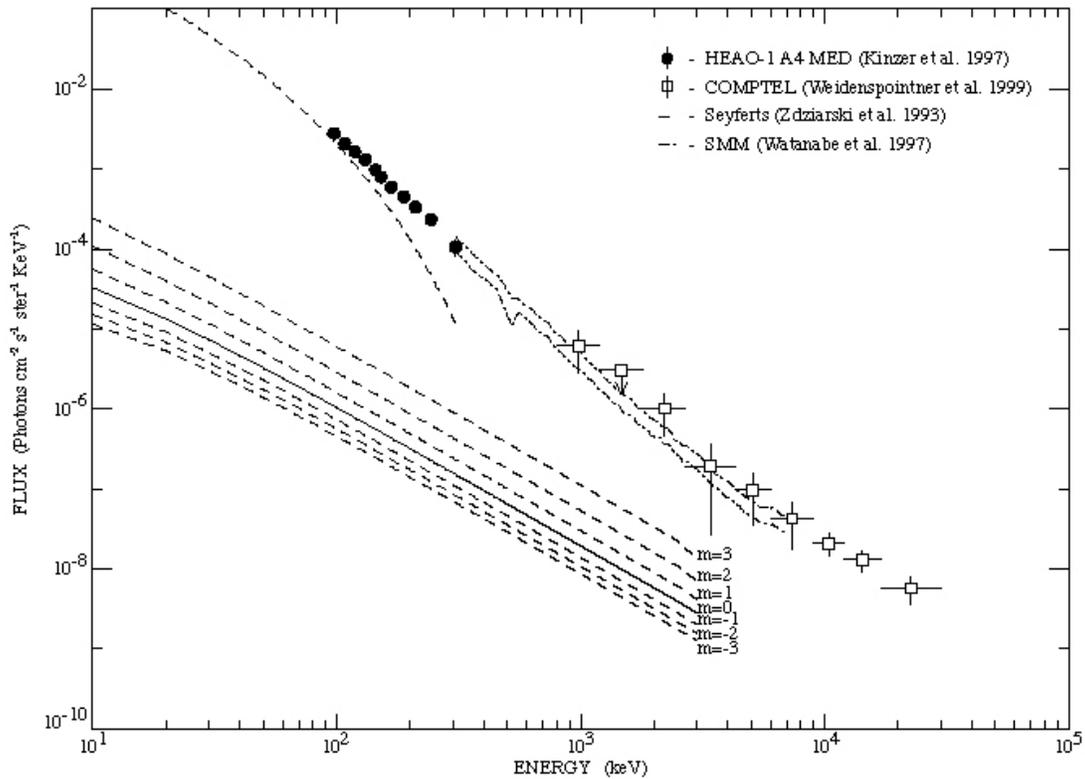

**FIGURE 1.** Estimated contribution of Cen A-like radio galaxies to the CGB for various assumptions about density evolution: $n(z) = n_0 (1+z)^m$ with $n_0 = 2.0 \times 10^{-6}$ galaxies Mpc$^{-3}$ and m = (3, 2, 1, 0, -1, -2, -3).

## Results

FIGURE 2 shows the contributions of GRBs and Cen A-like (FRI) radio galaxies to the CGB in comparison to contributions from SNIa and SNII (Watanabe et al. 1999a). The observed data are HEAO-1 (Kinzer et al.1997), COMPTEL (Weidenspointner 1999), and SMM (Watanabe et al. 1997). At low energies the contribution from Seyferts dominates (e.g., Zdziarski et al. 1993), while blazars dominate above a few MeV. Supernovae (mostly SNIa) dominate the MeV regime, where GRBs and Cen A-like radio galaxies (without density evolution) account for ~1% of the observed flux.

## Conclusion

While Seyfert galaxies and SNIa are the major contributors to the CGB in the 0.1-1 MeV band, GRBs and Cen A-like radio galaxies provide a non-negligible portion of the flux near 300 keV. However, the apparent flux deficit is not fully accounted for by the contributions from bursts and radio galaxies. The fact that a gap between Seyferts and SNIa remains suggests that there are contributing sources that have not yet been recognized. INTEGRAL might discover a new class of sources, but will also improve our understanding of radio galaxies and thereby improve existing models of the CGB.

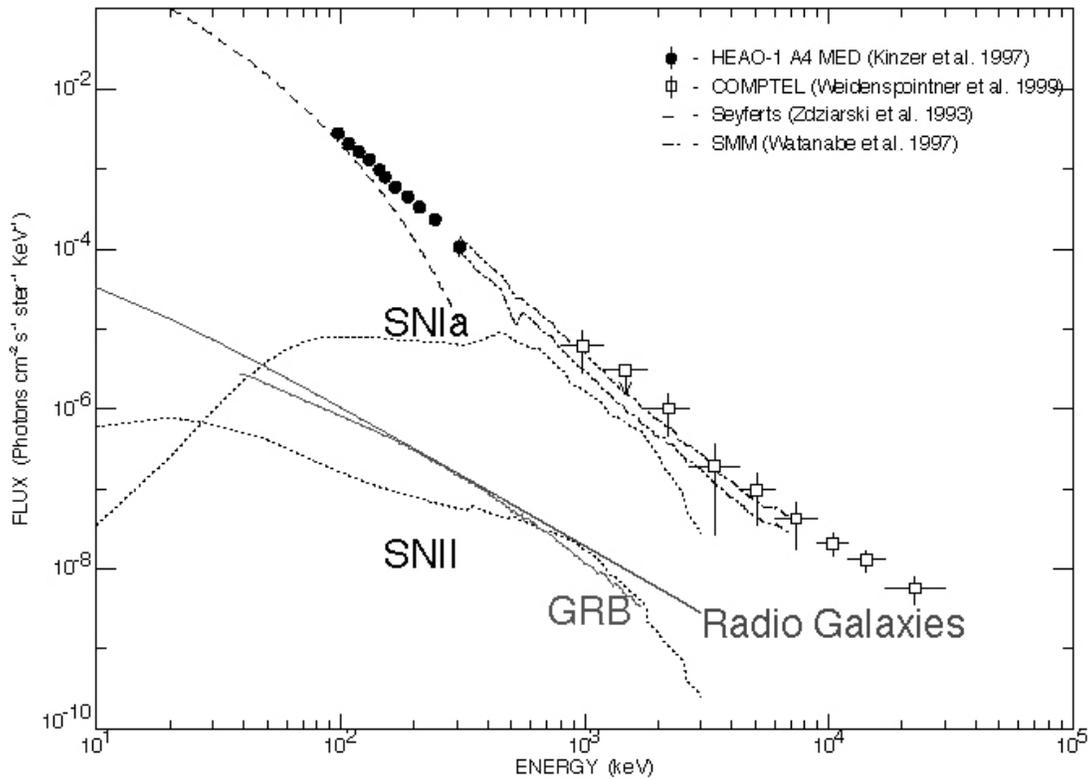

**FIGURE 2.** The observed CGB and estimated contributions from supernovae, gamma-ray bursts, and radio galaxies (without density evolution). While SNIa clearly account for the bulk of the observed CGB, up to 10% of the flux in the MeV regime could be due to SNII, GRBs, and radio galaxies.